\documentclass[openacc]{rstransa}%%%%where rstrans is the template name

\usepackage{aas_macros}

\newcommand{\Planck}{{\it Planck}}
\newcommand{\LCDM}{{$\Lambda$CDM}}
\newcommand{\Omegal}{{\Omega_{\Lambda}}}
\newcommand{\Omegam}{{\Omega_{m}}}
\newcommand{\seight}{\sigma_8}
\newcommand{\eightmpc}{8h^{-1} {\rm Mpc}}
\newcommand{\impc}{{\rm Mpc}^{-1}}

\newcommand{\kcmb}{\kappa_\mathrm{CMB}}

\newcommand{\Cl}{C_{L}}
\newcommand{\intp}{\int_0^{\infty}}
\newcommand{\ba}{\begin{eqnarray}}
\newcommand{\ea}{\end{eqnarray}}
\newcommand{\balpha}{\boldsymbol{\alpha}}

\def\n{{\bf n}}
\def\Cell{C_{\ell}}
\def\bell{{\boldsymbol{ \ell}}}
\def\bL{{\bf L}}

\titlehead{Research}

\begin{document}

%%%% Article title to be placed here
\title{Assessing the growth of structure over cosmic time with CMB lensing}

\author{%%%% Author details
  Mathew S. Madhavacheril$^{1}$}
%, X. Second author$^{2}$ and X. Third author$^{3}$

%%%%%%%%% Insert author address here
\address{$^{1}$Department of Physics and Astronomy, University of Pennsylvania, Philadelphia, PA 19104\\
}
%$^{2}$Second author address\\
%$^{3}$Third author address

%%%% Subject entries to be placed here %%%%
%\subject{xxxxx, xxxxx, xxxx}

%%%% Keyword entries to be placed here %%%%
%\keywords{xxxx, xxxx, xxxx}

%%%% Insert corresponding author and its email address}
\corres{Mathew S. Madhavacheril\\
\email{mathm@sas.upenn.edu}}

%%%% Abstract text to be placed here %%%%%%%%%%%%
\begin{abstract}
The standard \LCDM~ cosmological model informed by cosmic microwave background (CMB) anisotropies makes a precise prediction for the growth of matter density fluctuations over cosmic time on linear scales. A variety of cosmological observables offer independent and complementary ways of testing this prediction, but results have been mixed, with many constraints on the amplitude of structure $S_8$ being 2-3$\sigma$ lower than the expectation from \Planck~ primary CMB anisotropies. It is currently unclear whether these discrepancies are due to observational systematics, non-linearities and baryonic effects or new physics. We review how gravitational lensing of the CMB has and will continue to provide insights into this problem, including through tomographic cross-correlations with galaxy surveys over cosmic time. 
\end{abstract}
%%%%%%%%%%%%%%%%%%%%%%%%%%%

%%%%%%%%%% Insert the texts which can accomdate on firstpage in the tag "fmtext" %%%%%
%\begin{fmtext}

\maketitle
%\end{fmtext}

%%%%%%%%%%%%%%% End of first page %%%%%%%%%%%%%%%%%%%%%

\section{Introduction}

The cosmic microwave background (CMB) is a key prediction of the hot big bang and consists of photons that mostly last scattered during the epoch of recombination at around $t=380,000$ years. Measurements of CMB anisotropies by the {\it COBE} satellite \cite{astro-ph/9401012}  provided our first view of fluctuations in the early universe, providing a handle on physics from well before the epoch of structure formation. Subsequent measurements by the {\it WMAP} satellite \cite{astro-ph/0302209,1212.5226} (together with complementary balloon and ground-based experiments \cite{astro-ph/9905100,astro-ph/9911445,astro-ph/0005123}) ushered in an era of precision cosmology, cementing the $\Lambda$-Cold Dark Matter (\LCDM) model. With state-of-the-art primary CMB measurements from the {\Planck} satellite \cite{1303.5076,1502.01589,1807.06209,1807.06205,10.1093/mnras/stac2744,10.1051/0004-6361/202348015}, the community now has a precise benchmark model to compare other observations to. Increasingly precise measurements of the CMB damping tail and CMB polarization by high-resolution ground-based experiments like the South Pole Telescope (SPT; e.g \cite{2212.05642}) and the Atacama Cosmology Telescope (ACT; e.g. \cite{2007.07288,2007.07289}) are now on a path to enhancing the \Planck~ legacy, especially for extensions to the \LCDM~ model (e.g. \cite{2109.04451}).

At the same time, multiple distinct probes of the late-time universe continue to offer a test of the standard model, including galaxy redshift space distortions (RSD) \cite{1909.05277,1909.05271,2007.08991}, the Lyman-$\alpha$ forest \cite{2007.08991,2309.03943,2403.08241}, the abundance of galaxy clusters \cite{2401.02075,2402.08458} and galaxy weak lensing \cite{2105.13549,2107.00136,2304.00701,2304.00705,2304.00704,2007.15632}.  Secondary anisotropies in the CMB itself have become an informative source of information on large-scale structure through the variety of ways in which CMB photons interact with matter since the recombination epoch. This includes measurements of gas pressure and density through the thermal and kinetic Sunyaev-Zeldovich (SZ) effects \cite{10.1146/annurev.astro.40.060401.093803,1811.02310,1810.13423,1810.13424,1901.02418} respectively.  Here, we review the status of weak gravitational lensing of the CMB\cite{astro-ph/0601594} as a probe of the growth of structure.

The CMB allows for a semi-model-independent test of the standard model. Providing a snapshot of the $z\approx 1090$ ($t\sim 380,000$ yr) universe that can be modeled precisely with linear physics, it allows for precise constraints on parameters of the standard \LCDM~ model (with the \Planck~ satellite providing the state-of-the-art constraints today).  This fit model can then be extrapolated to late times, with the growth of density perturbations on large scales modeled precisely with linear theory. A useful proxy for describing growth is the number $\sigma_8$, which measures the root-mean-square (RMS) amplitude of fluctuations in the total matter density assuming linear theory. The amplitude of fluctuations increases with cosmic time (or decreasing redshift $z$) through $\sigma_8(z)=D(z)/D(0)\sigma_8$ due to the growth of perturbations under gravity (elaborated more in Sec \ref{sec:growth}).   Some observables are sensitive to the combination $S_8 \equiv \sigma_8 (\Omegam / 0.3)^{0.5}$, where $\Omegam$ is the total matter density.  As suggested in \cite{astro-ph/9611077}, with the CMB prediction in hand, the amplitude of structure can then be measured more directly with late-time growth probes, although in many cases, significant modeling of non-linear physics is required. 

Many probes of the late-time universe are in mild disagreement with the \Planck~ prediction. Disagreements between the early universe prediction and the late universe may either be to systematics in either CMB or late-time measurements, inadequacies of modeling of non-linear physics or new physics. We will discuss in this review how CMB lensing has recently been used to distinguish between systematics and new physics by providing an accurate view of the matter power spectrum primarily at wave-numbers $k<0.1 \impc$ and redshifts $z=1-5$.

In Sec \ref{sec:growth}, we review the physics behind growth of structure in the standard model, while in Sec \ref{sec:lensing} we explore how CMB lensing serves as a probe of growth. We conclude placing recent CMB lensing measurements in the broader context of other large-scale structure probes in \ref{sec:broader}.

\section{Growth of structure}\label{sec:growth}

Primordial density fluctuations grow under gravity as the universe expands.  General relativity applied to a universe with a few components (cold dark matter, baryons, photons, neutrinos and dark energy) produces precise predictions on linear scales for how structure grew from the initial perturbations.  We review here how the standard cosmological model allows for this prediction.

The amplitude $A_s$ and spectral tilt $n_s$ of the initial perturbations may have been set through a process like cosmic inflation, which naturally predicts a nearly scale-invariant slightly red spectrum.  The primordial scalar amplitude and spectral tilt are well constrained by observations of the CMB anisotropies \cite{Planck2018legacy}.  The subsequent growth of perturbations depends on scale and time. Since the horizon expands in a post-inflationary expanding universe, all modes of relevance today were once super-horizon and subsequently enter the horizon at some epoch determined by the mode wavenumber $k=|{\bf k}|$. Relativistic perturbation theory predicts that super-horizon matter overdensities $\delta \equiv (\rho-\bar{\rho})/\bar{\rho}$ scale as  $\delta\propto a^2$ during radiation domination and $\delta\propto a$ during matter domination, where $a$ is the scale factor, $\rho$ is the total matter density and $\bar{\rho}$ is the mean total matter density.   As the universe transitions from radiation through matter domination to eventual dark energy domination, sub-horizon perturbations, on the other hand, grow slowly ($\delta\propto {\rm log} ~a$), quickly ($\delta \propto a$) and not at all ($\delta = {\rm const.}$) respectively in each consecutive epoch. The largest observable scales ($k<k_{\rm eq}$)  today never experienced slow growth during radiation domination (since they entered the horizon relatively recently) and hence retain the nearly-scale invariant spectrum of the primordial perturbations.  Smaller scales spent more time with logarithmic growth and hence are suppressed leading to a characteristic peak in the matter density power spectrum at the equality scale $k_{\rm eq}$.  

To predict the matter power spectrum at late times, in many models\footnote{Models with massive neutrinos and modified gravity are notable exceptions, where the transfer function becomes redshift dependent or equivalently the growth factor becomes scale dependent.}, it is possible to factor out the time-dependent growth $\delta({\bf k},a) \propto D(a)\delta({\bf k},1)$ from the scale-dependent transfer function $T(k)$ that transforms the initial primordial fluctuations $\Phi({\bf k})$ through the above physics into the post-recombination epoch.  These considerations lead to a prediction for the linear matter power spectrum :

\begin{equation}
    \label{eq:plin}
k^3 P^L_{mm}(k,z) = \frac{8\pi^2}{25} \frac{k^4}{\left(\Omega_m H_0^2\right)^2} A_s \left(\frac{k}{k_p}\right)^{n_s-1} T(k)^2 D(a)^2
\end{equation}

with redshift $z$, mean matter density parameter today $\Omegam$, Hubble expansion rate today $H_0$ and  pivot scale $k_p$. In a matter-only Einstein-de-Sitter (EdS) universe (dark energy density $\Omega_{\Lambda}=0$), the growth function can be analytically obtained to be $D(a)=a$.\footnote{Note that many authors (e.g. \cite{2212.05003}) differentiate between the growth factor $D(a)$ and the growth suppression factor $g(a)$ such that $D(a)=ag(a)/g(0)$, where the latter is always normalized to unity today with $D(1)=1$. With this alternative definition, $g(1)$ depends on the cosmological model and the power spectrum ends up proportional to $(ag(a))^2$.}  A closed form solution is possible in a flat \LCDM~ universe too (albeit with constant dark energy equation of state),   in terms of a hypergeometric function (see Appendix \ref{app:dgrowth}). 

\begin{equation}
\label{eq:dgrowth}
\frac{D(a)}{a} = \sqrt{1+x^3} ~~{}_{2}F_{1}(\frac{5}{6},\frac{3}{2},\frac{11}{6},-x^3)
\end{equation}
\begin{equation}
x = \frac{\Omega_\Lambda}{\Omega_m} a
\end{equation}

\begin{figure}
\includegraphics[width=\columnwidth]{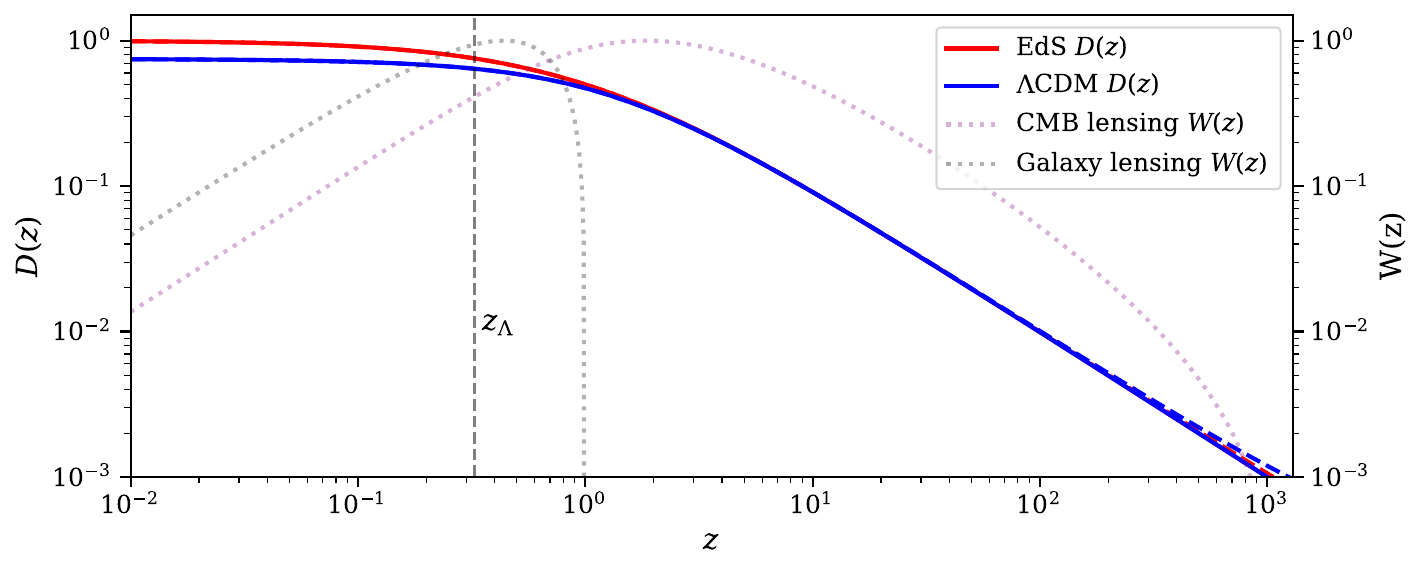}
  \caption{Linear growth factors $D(z)$ in an EdS and \LCDM~ universe are shown in red and blue, respectively.  The solid lines are the analytic form in Eq. \ref{eq:dgrowth} and the dashed lines are an exact numerical calculation from \texttt{CLASS}; the former ignores radiation. The dotted lines show lensing kernels $W(z)$ for a galaxy lensing source at $z=1$ (light gray) and for CMB lensing (light purple). The vertical dashed line marked $z_{\Lambda}$ marks the redshift at which the matter density begins to drop below the dark energy density causing the cosmological constant to begin dominating the dynamics of the universe. }
    \label{fig:dgrowth}
   
\end{figure}

The latter formula is useful for quick closed-form evaluation of the linear matter power (when combined with empirical fitting formulas for the transfer function, e.g. \cite{Eisenstein, EisensteinHu}), though it becomes inaccurate (>5\% error) at high redshifts ($z>200$) since radiation is neglected. In practice, predictions for the linear matter power spectrum can be obtained with high accuracy with Einstein-Boltzmann codes \cite{CAMB,class} that solve the linear-order differential equations involved in the Boltzmann hierarchy of all the components, and reliably accelerated with emulators \cite{2106.03846,2303.01591,2405.07903}.

We will discuss many observables that are built from $P^L_{mm}(k,z)$\footnote{Or when studying galaxy clustering, from related quantities such as auto and cross-correlations of CDM and baryons.}, though in general, they all introduce (to different extents) (a) dependence on higher moments of $P^L$ either due to non-linear evolution or because the observable goes beyond two-point statistics (e.g. cluster abundances, bispectra, PDFs, peak counts) and (b) integration over redshifts due to projection along the line-of-sight or due to uncertain redshift information. Since these complications hide the underlying $P^L_{mm}(k,z)$ object, it is customary to instead report a proxy for its normalization. Historically, this has been the $\sigma_8$ parameter, defined as the root-mean-square (RMS) amplitude of the linear matter density field today when smoothed with a top-hat filter\footnote{The motivation for this quantity stems from its importance in the characterization of halos in the Press-Schechter \cite{PressSchechter} formalism and its derivatives.} with radius $R=\eightmpc$

\begin{equation}
\label{eq:sigma2}
\sigma^2(R,z) = \int_0^{\infty} d{\rm ln}k \frac{k^3}{2\pi^2}P^L_{mm}(k,z)W^2(kR)    
\end{equation}

with $\sigma_8\equiv \sigma(R=\eightmpc,z=0)$ and the Fourier transform of the top-hat is given by

$$
W(kR) = 3 \frac{{\rm sin}(kR)-kR~{\rm cos}(kR)}{\left(kR\right)^3}
$$

which asymptotes to unity on large scales $k\rightarrow0$ as expected for a smoothing filter.  Eq. \ref{eq:sigma2} should be understood as summing up variance contributions from each wave-number, with $P(k,z)$ being the variance per mode, the number of modes in each Fourier shell proportional to $k^2 dk$ and $W(kR)$ being the field-level smoothing filter. It can alternatively also be read as the sum of contributions of the dimensionless filtered power spectrum $(k^3P^L_{mm}(k)/2\pi^2)  W^2$ per unit decade $d{\rm ln}k$. With the $R=\eightmpc$ smoothing, $\seight$ converges in a typical \LCDM\ model at an upper cutoff of $k_{\rm max}\approx 0.2 \impc$  in the integral in Eq. \ref{eq:sigma2}, whereas the variance never converges without smoothing ($W=1$).  In models where the redshift evolution can be factorized out, we have from Eq \ref{eq:plin} and Eq \ref{eq:sigma2}, 

$$
\sigma_8(z)=\frac{D(z)}{D(0)}\sigma_8.
$$

Several caveats regarding constraints on $\seight$ (and related quantities such as $S_8 \equiv \seight (\Omega_m/0.3)^{0.5}$) are worth noting given that it involves an extrapolation of the linear matter power:
\begin{enumerate}
  \item It is a ``derived parameter'' computed based on a prediction for $P^L_{mm}(k,z=0)$.  The amplitude of structure at the present day is not a free parameter in standard cosmology (nor is it in fact directly observable since the available cosmological volume precisely at $z=0$ vanishes). In a typical analysis, Einstein-Boltzmann codes that take as input the amplitude of the initial perturbations $A_s$ are used to predict functions of $P^L_{mm}(k,z)$ that are fit to data and then subsequently extrapolated to $z=0$.  Many analyses sample in $A_s$ (or its logarithm), while some that sample $\sigma_8$ or $S_8$ determine $A_s$ through `shooting` methods. While $\seight$ scales straightforwardly with $A_s$ as $\propto \sqrt{A_s}$, its scaling with $\Omega_m h^2$, for example, is less intuitive: $\seight$ generally increases with the physical matter density $\Omega_m h^2$ even though the peak amplitude of the power spectrum is reduced because the equality scale $k_{\rm eq}$ in $T(k)$ is pushed to higher wavenumbers (since matter-radiation equality happens earlier) increasing the overall variance in the field.
    \item The linear matter field appears in the definition of $\seight$, but any observable with an arbitrary non-linear dependence on the matter field can be and often is used to constrain it.
    \item While the predicted value of $\seight$ is only sensitive to $P^L_{mm}(k<0.2 \impc,z=0)$ during modeling, observables that do not measure those scales (for example, with sensitivity only to $k\sim 2 \impc$) and observables that measure completely unrelated epochs ($z>0$) can and often are used to constrain it. The provided model is used to extrapolate to the relevant scales and the present day.
\end{enumerate}

As an exercise in understanding the consequences of the above, consider an imaginary model `aCDM', imagine that all parameters except $A_s$ are known to high precision and ignore non-linearities. The model `aCDM' is identical to \LCDM\ but predicts (with no free parameters) that the linear matter power spectrum should be suppressed by a factor of 10 (relative to the \LCDM\ prediction)  between $0.4 < z < 0.5$ only on scales $1 \impc < k < 3 \impc$.  If using observables that skillfully avoid this redshift and wavenumber range, then both models will give identical fits for $A_s$ and hence $\seight$. If one uses an observable that {\it only} uses this particular range of scales and redshifts, then the \LCDM\ fit would produce a low $\seight$, but the aCDM fit would produce the higher value obtained by datasets that avoid this redshift and scale range.

Today, parameters such as the amplitude of scalar perturbations $A_s$, their spectral tilt $n_s$, the physical CDM density $\Omega_c h^2$, the physical baryon density $\Omega_b h^2$ and the combination $\Omega_m h^3$ are determined\footnote{$\Omega_m h^3$ is the parameter combination constrained by the very well measured angular scale of the sound horizon $\theta_{*}$, which together with other parameters constrains the expansion rate today $H_0$. Note that $h \equiv H_0$ / (100 km/s/Mpc) throughout this article.} to very high precision through measurements of the CMB anisotropies that probe the matter field at $z\sim 1090$ through its effect on radiation. This provides a prediction for $\seight$ extrapolated over more than 13.3 billion years with Eq \ref{eq:plin}.  Late-time observables discussed later in this article probe the matter field at more recent times. When $\seight$ measurements from the latter disagree with the CMB prediction assuming \LCDM, they provide a useful diagnostic for both our understanding of the growth of structure and non-linear and baryonic effects that complicate the mapping between observables and $P^L_{mm}$.

Many probes are sensitive to the velocity field, rather than or in addition to the density field.  In linear theory, velocities are related to the underlying density through the continuity equation, which in Fourier space is given by ${\boldsymbol{v} } = (ifaH/k) \hat{\boldsymbol{k}}~ \delta_m$, for scale factor $a$, Hubble expansion rate $H(a)$, comoving wave-number $\vec{\boldsymbol{k}}$ and with the growth {\it rate} defined by 
$$
f = \frac{d{\rm ln} D}{d{\rm ln} a}
$$

Galaxy clustering in particular picks up a dependence on the velocity field.  Redshift space distortions (RSD) are anisotropic distortions in the power spectrum of galaxies due to peculiar velocity contributions to measured redshifts. These measurements are sensitive to the parameter combination $f\seight$.  An alternative approach for constraining $f\seight$ feasible in the local universe is to use peculiar velocities directly measured through distance indicators \cite{1912.09383}. The kinetic Sunyaev-Zeldovich (kSZ) effect in CMB measurements is also sensitive to $f\seight$ since kSZ secondary anisotropies are proportional to peculiar velocities of ionized electrons \cite{1408.6248,2101.08374}. However, this cosmological information is perfectly degenerate with the optical depth of galaxies \cite{1810.13423,1901.02418}.

\section{Weak lensing as a probe of growth}\label{sec:lensing}

We will now specialize to the use of weak lensing as a probe of growth, with particular focus on CMB lensing. Gravitational weak lensing observables probe the line-of-sight integrated matter content in the universe. Background light sources are distorted by deflection in the gravitational potentials between the light source and the observer. CMB photons in particular are sourced from the recombination epoch with $z\approx 1090$ and therefore undergo deflections throughout the history of structure formation.  The sensitivity of weak lensing to the line-of-sight integrated matter overdensity $\delta_m$ is often characterized by the lensing convergence field (a function of angular direction $\mathrm{\bf{\hat{n}}}$) \cite{astro-ph/9912508}

\begin{equation}
  \label{eq:kappa}
    \kappa(\mathrm{\bf{\hat{n}}}) = \int_0^{\infty} dz W^\kappa(z) \\
                \delta_{m}(\chi(z)\mathrm{\bf{\hat{n}}}, z).
\end{equation}

Here, the lensing kernel that determines weights from different epochs is

\begin{equation}
  \label{eq:kernel}
    W^\kappa(z) = \frac{3}{2}\Omega_{m} H_0^2 \\
            \frac{(1+z)}{H(z)} \frac{\chi(z)}{c} \int_z^{\infty} dz' n(z') \\
                \frac{\chi(z') - \chi(z)}{\chi(z')},
\end{equation}

where $\chi(z)$ is the comoving distance to redshift $z$, $H(z)$ is the Hubble expansion rate and $c$ is the speed of light.
Since the CMB lensing source redshift $n(z)$ can be approximated by a Dirac delta function at the recombination redshift $z_\star$, this reduces to

\begin{equation}
    W^{\kcmb}(z) =  \frac{3}{2}\Omega_{m}H_0^2  \frac{(1+z)}{H(z)} \frac{\chi(z)}{c} \\ 
    \left [ \frac{\chi(z_\star)-\chi(z)}{\chi(z_\star)} \right ].
    \label{eqn:cmb_kernel}
\end{equation}

The convergence is proportional to the line-of-sight integrated mass density and is related to the 2d potential $\phi$ and deflection angle $\alpha$ through 

\ba
\nabla^2 \phi = {\boldsymbol{\nabla}}.{\boldsymbol \alpha} = 2 \kappa.
   \label{eqn:lensing}
\ea

The lensing kernels for CMB lensing (with source redshift at $z=1090$) and a representative galaxy weak lensing sample at $z=1$ are shown in Fig. \ref{fig:dgrowth}.  CMB lensing observables have a broad kernel with sensitivity to all redshifts across the epoch of structure formation, with contributions peaking at $z\sim 1-3$ and significant contributions at $z>5$.

A general expression for the angular power spectrum of two fields $a$ and $b$ under the Limber approximation is
\ba
\label{eq:clab}
\Cl^{a,b} = \intp dz \frac{H(z)}{c\chi^2} W^a(z)W^b(z) P^{\rm NL}_{mm}(z,k=L/\chi(z))
\ea

where $W$ could either be a lensing kernel $W^{\kappa}$ from Eq. \ref{eq:kernel} when a lensing convergence field is involved or the redshift distribution $W^g\equiv b_g dn/dz$ of a galaxy tracer with linear bias. Various observables can be calculated with this expression:  
\begin{enumerate}
\item {\bf CMB lensing auto-spectrum: } When $W^a=W^b=W^{\kcmb}$, we obtain an expression for the CMB lensing auto-spectrum. Accurately estimating this power spectrum requires a careful understanding of several complexities (see Sec \ref{sec:challenges}) including instrumental noise and systematics, astrophysical foregrounds and higher-order contributions.  With these concerns mitigated using improvements to the reconstruction algorithm, if the non-linear matter power spectrum  that appears in Eq. \ref{eq:clab} is known exactly (given a set of cosmological parameters), then there are typically no additional nuisance parameters\footnote{Some analyses, e.g. \cite{SPT}, have subtracted or marginalized over a foreground bias template. Some authors (e.g. \cite{2103.05582}) have also advocated for the marginalization of baryonic feedback parameters for future analyses, although the same work suggests alternative techniques that do not require marginalization.}. 
\item {\bf Cosmic shear:} When $W^a=W^b=W^{\kappa,\rm gal}$ where the lensing kernel in Eq. \ref{eq:kernel} includes a redshift distribution function $n(z)$ for the source galaxies. Several sources of additional uncertainty are introduced, including shear multiplicative biases (e.g \cite{astro-ph/0506030,astro-ph/0506112} or the more recent exploration in \cite{10.3847/1538-4357/abb595}), photometric redshift uncertainties (e.g. \cite{astro-ph/0003380,2206.13633}) and intrinsic alignments (e.g. \cite{1708.09247} and more recently \cite{2309.16761}), which typically introduce additional nuisance parameters for each redshift bin.  In addition, since the sources are at much lower redshift than CMB lensing, typical angular scales correspond to higher wavenumber $k$, making this observable more sensitive to non-linear evolution and baryonic feedback. On the other hand, cosmic shear has significantly more weight around the dark-energy dominated epoch (see Fig \ref{fig:dgrowth}), allowing for complementary exploration of modified gravity and dark energy phenomenology.
\item {\bf Galaxy-galaxy lensing and CMB lensing tomography: }  When one field is a weak lensing observable $W^a=W^{\kappa,\rm gal}$ or $W^a=W^{\kappa,\rm CMB}$  and the other is a biased galaxy tracer overdensity $W_b=b_g(dn/dz)$ with linear galaxy bias $b_g$ and redshift distribution $dn/dz$, the cross-correlation $C_{\ell}^{\kappa g}$ is sensitive to $b_g \sigma_8^2$. This is often combined with the angular clustering of galaxies itself $C_{\ell}^{gg}$ with $W_a=W_b=b_g(dn/dz)$, which is sensitive to $b_g^2 \sigma_8^2$. The joint analysis of these two observables (sometimes referred to as a 2x2 analysis \footnote{The Dark Energy Survey collaboration has popularized the $n$x2 terminology, which refers to a joint analysis of $n$ 2-point functions. 3x2 typically adds the auto-spectrum of the lensing field. Joint analyses featuring combinations with CMB lensing have produced 5x2 and 6x2 \cite{2206.10824} analyses to date.  }) allows the degeneracy between galaxy bias and $\sigma_8$ to be broken, providing an alternative channel for growth measurements.  These measurements require additional modeling beyond what is shown in Eq. \ref{eq:clab} for effects such as magnification bias and non-linear biasing. If the galaxy sample that is used is not a spectroscopic sample, additional nuisance parameters may be introduced to characterize its redshift distribution, or it may be calibrated with spectroscopic surveys through clustering redshifts (\cite{0805.1409} as done in e.g. \cite{unwise}). If the lensing information is obtained from galaxy sources as opposed to the CMB, the nuisance parameters involved in cosmic shear mentioned previously will also be needed.
\item {\bf CMB lensing-shear cross-correlation}: When one field is a galaxy weak lensing observable $W^a=W^{\kappa,\rm gal}$ and the other CMB lensing $W^b=W^{\kappa,\rm CMB}$, the resulting cross correlation is proportional to $\sigma_8^2$, providing yet another channel for structure growth constraints (e.g. \cite{1311.6200,2015PhRvD..92f3517L,2016MNRAS.459...21K,2019PhRvD.100d3501O,10.1093/mnras/stw947,2020ApJ...904..182M,2011.11613,2023PhRvD.107b3530C,2309.04412}). Once again, nuisance parameters involved in cosmic shear such as multiplicative shear biases, photometric redshift distribution uncertainty and intrinsic alignments will need to be marginalized over. However, including these cross-correlations in a 5x2 or 6x2 joint analysis can lead to improved mitigation of these systematics \cite{des}.
  
\end{enumerate}

\begin{figure}
\includegraphics[width=\columnwidth]{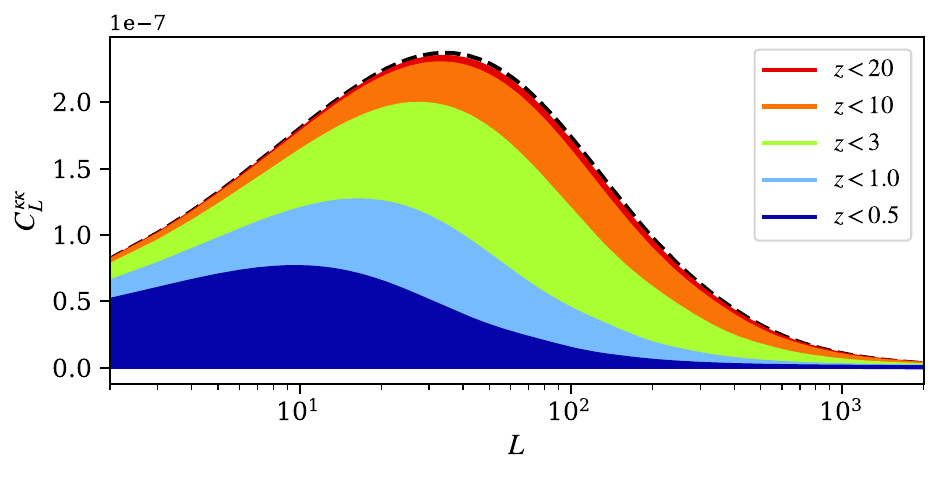}
  \caption{The CMB lensing auto-spectrum is shown here for cumulatively higher redshift cutoffs, from blue ($z<0.5$) to red ($z<20$) with the latter containing nearly all of the full CMB lensing power (dashed black line). Lower multipoles contain more information from lower redshifts with the angular power spectrum peak $L_p = k_{\rm eq} \chi(z_{\rm eff})$ shifting to higher multipoles as higher redshifts are included.}
    \label{fig:zdep}
   
\end{figure}

All analyses of (a) have been in harmonic space, whereas (b) and (c) have been measured in both harmonic and real-space, with the latter correlation functions calculated through a Hankel transform of Eq. \ref{eq:clab}.  Noise in the galaxy overdensity field can be roughly modeled with a shot noise contribution $N_{\ell} = 1/n_g$ where $n_g$ is the 2d number density of galaxies, while noise in the galaxy lensing field is from shape noise $N_{\ell} = \sigma_e^2/n_g$, where $\sigma_e^2$ is the variance due to both the intrinsic ellipticity distribution of galaxies and shape measurement noise (with typical values of around $\sigma_e\approx0.3$).    Current generation galaxy weak lensing surveys include the Dark Energy Survey (DES), Hyper Suprime Cam (HSC) and the Kilo-Degree Survey (KiDS). With their most recent data releases DES-Y3, HSC-Y3 and KiDS-1000, they have covered approximately 4000 sq. deg., 400 sq. deg. and 1000 sq. deg with effective source galaxy number densities of 5.6 gal/arcmin$^2$ \cite{2011.03408}, 19.9 gal/arcmin$^2$ \cite{2107.00136} and 6.2 gal/arcmin \cite{2011.03408}, respectively. Noise in the CMB lensing field will be discussed in the next section.

In Fig. \ref{fig:zdep}, we show how various redshifts cumulatively contribute to the CMB lensing signal evaluated through Eq. \ref{eq:clab}. Structures at lower redshifts and more recent times contribute more on large scales or low multipoles. A large fraction of the $L<20$ power is contributed by $z<0.5$ structures whereas most of the power is from higher redshifts for multipoles $L>100$.  The shape of the CMB lensing auto-spectrum therefore contains some information on $\sigma_8(z)$ and future CMB surveys may be able to perform tomography to some degree with just the auto-spectrum.

\subsection{Measuring CMB lensing}

CMB experiments measure the temperature anisotropy $T$ and Stokes components $Q$ and $U$ of its polarization anisotropies. The observed lensed components are related to the unlensed sky and the deflection angle $\balpha$ by

\ba
\label{eq:remap}
T(\n)&=&\bar{T}(\n + \balpha) \\
Q(\n)&=&\bar{Q}(\n + \balpha) \nonumber \\
U(\n)&=&\bar{U}(\n + \balpha) \nonumber
\ea

The deflection field $\balpha$ is related to the 2d lensing potential $\phi$ and lensing convergence $\kappa$ through Eq. \ref{eqn:lensing}, with the latter proportional to a weighted line-of-sight integral of the matter density all the way to the redshift of last scattering $z_{\star}\approx 1090$.  The resulting deflection angles are typically of the order of arcminutes, but these deflections are coherent over degree scales corresponding to the size of typical lenses projected along the line-of-sight.

Algorithms for measuring the CMB lensing signal aim to reconstruct maps of the lensing convergence $\hat{\kappa}(\n)$ from observed maps  $X(\n)$ of the CMB temperature and polarization (containing instrument beam, noise and astrophysical foregrounds).  The maps can then be subsequently used for cross-correlations with other observables or they can be used to estimate the CMB lensing auto-spectrum $C_L^{\kappa \kappa}$. Most CMB lensing analyses to date have used `quadratic estimators' (QEs) \cite{HuOk} to perform this reconstruction \footnote{A Bayesian approach has been used by SPT in \cite{2012.01709} and iterative methods have been proposed in \cite{1704.08230}.}. These exploit the fact that the primary (unlensed) CMB is statistically isotropic and captured fully by the power spectra $\Cell^{TT}$, $\Cell^{TE}$, $\Cell^{EE}$ of the CMB temperature and E-mode transform of the polarization signal, with no coupling between the different $\bell$ modes.   However, the lensing remapping in Eq. \ref{eq:remap} introduces mode-coupling whose expectation value (to leading order) is proportional to modes $\kappa({\bf L})$ of the lensing convergence. This motivates estimators formed from weighted sums of quadratic pairs of CMB temperature and polarization fields. 

The resulting reconstruction $\hat{\kappa}(\bL)$ (the inverse harmonic transform of the map) has a noise spectrum characterized by $N_L^{\kappa \kappa}$. This contains contributions not just from the instrument beam and noise but also the signal power in the CMB, since chance correlations in the primary unlensed CMB can be confused with lensing. Reconstruction on large scales (small $L$) depends primarily on small-scales in the input CMB maps (high $\ell$); in fact, ground-based experiments like ACT and SPT now measure lensing modes on scales much larger than what they reliably observe the CMB itself on.  These CMB lensing maps (with small well-understood Monte Carlo corrections to their overall normalization) have been produced by \Planck \cite{1502.01591,1807.06210,2206.07773}, ACT \cite{1611.09753,2004.01139,2304.05203,2304.05202} and SPT \cite{2308.11608,1905.05777,1705.00743} for cross-correlation with external tracers.

Significant additional work is needed to transform the reconstructed CMB lensing map to an estimate of its auto-spectrum.  Since the lensing reconstruction is a quadratic estimate of the CMB maps, the auto-spectrum of this map is a 4-point function.  The disconnected or Gaussian part of this 4-point function is non-zero even in the absence of lensing; it is much larger than the connected signal from lensing itself and needs to be subtracted.  We discuss this briefly in the next section. Several other biases such as the mean-field (due to mask or noise-induced anisotropy, see e.g \cite{1209.0091}) and an $N_1$ bias due to secondary contractions of the trispectrum \cite{astro-ph/0302536,1008.4403} are also estimated and subtracted.

\subsection{Challenges for CMB lensing probes}
\label{sec:challenges}

{\bf Foregrounds: } Since ground-based surveys began to make strides in CMB lensing measurements post-\Planck, astrophysical foregrounds have become one of the main systematic effects investigated.   Following the first correlated extra-galactic simulation suite targeting CMB experiments \cite{Sehgal}, the authors of \cite{vanEngelen2014} demonstrated how arcminute-resolution experiments like ACT and SPT are susceptible to large biases in CMB lensing reconstruction due to tSZ and CIB contamination.  These biases arise mainly due to non-Gaussianity of the foregrounds themselves, whose large-scale structure bispectra and trispectra between themselves and the CMB lensing field itself appear in both the CMB lensing auto-spectrum as well as cross-correlations of the CMB lensing map with LSS tracers.  

Rather than modeling the complicated astrophysics behind these non-Gaussian biases, the community has oriented towards developing methods to fully mitigate them. Many new modifications to the standard QE have since been proposed including (a) a geometric method involving deprojection of mode-coupling introduced by point source and cluster-like objects \cite{Osborne,Sailer} (b) isolating the shear component of CMB lensing which is generally foreground-immune \cite{SchaanFerraro,QuShear} and (c) using foreground-cleaning selectively in parts of the QE \cite{MMHill,DarwishEtAl,2012.04032}. The above methods have been demonstrated to be sufficient to fully mitigate foreground biases with minimal loss of SNR for current surveys like ACT and SPT \cite{MacCrannForegrounds}. They can also be combined optimally \cite{2108.01663,2111.00462} allowing future surveys like the Simons Observatory to remain robust. Beyond SO (below white noise levels of $5 \mu K-$arcmin), the noise levels improve to a level such that most of the lensing information is derived from the polarization information. Since extra-galactic foregrounds are expected to not be significantly polarized, foreground biases become much less of a concern (as is already the case with deep CMB experiments like SPT-3G \cite{SPT3glensing}).

{\bf Instrumental systematics and noise modeling: } The auto-spectrum of the CMB lensing map is a 4-point function that has large contributions from the disconnected Gaussian component. This component depends on both the true signal 2-point power of the CMB as well as foreground and noise power.  The noise power in particular is hard to model for ground-based experiments with complicated scan patterns and large atmospheric noise contributions, although remarkable progress has been made in recent years \cite{AtkinsNoise}.  Since \cite{Namikawa}, many CMB lensing analyses (including by the \Planck~ collaboration) have used a `realization-dependent' simulation approach that uses combinations of the realized data and Monte Carlo simulations to estimate this noise bias in a way that is not sensitive to mis-matches between the data and simulation power spectra to linear order. This technique proved insufficient for ACT, which instead used \cite{2304.05202} a newly developed cross-estimator technique from \cite{MMNoise}.  That work developed an efficient algorithm that can use four or more splits of the data with independent noise (e.g. collected at different times) such that no split is ever repeated in the lensing 4-point estimator. This results in $N_0$ and mean-field biases that are completely unaffected by assumptions in the instrument noise modeling.

Many instrumental systematics such as polarization angle mis-calibration, beam asymmetries, boresight pointing errors, gain drifts, gain calibration errors and electrical crosstalk \cite{2311.05793} can potentially introduce statistical anisotropy that mimics lensing. A study from \cite{2011.13910} found that for an SO-like experiment with expected levels of these systematics, the bias to lensing reconstruction is less than $0.6 \sigma$.  However, further work is needed to test these systematics in the presence of atmospheric noise and realistic scan strategies.  In its recent high-significance ACT DR6 lensing auto-power spectrum analysis, the authors implemented conservative toy models for a number of systematics including gain miscalibration, beam uncertainty, temperature-to-polarization leakage, polarization efficiency and angle mis-calibration and found that none of these resulted in $>0.4\sigma$ biases in the baseline analysis.  In addition, more than 200 null tests sensitive to various systematics were carried out, with no statistically significant failures seen. The ACT DR6 approach employed a blinding policy where cosmological inferences were carried out only after confirmation that the null test suite passed.

\begin{figure}
\centering
\includegraphics[width=0.7 \columnwidth]{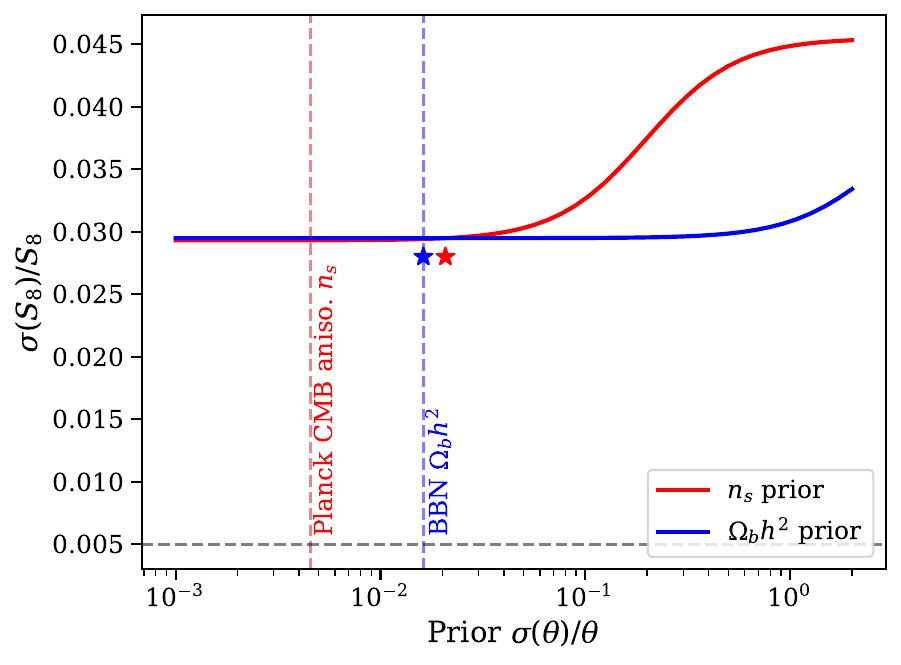}
  \caption{CMB lensing analyses typically use a prior on the spectral index and physical baryon density for `lensing alone' constraints. This information matrix calculation (loosely based on the experimental configuration in \cite{MM2023})  shows the relative uncertainty on $S_8$ as a function of the prior width on $n_s$ (red) and $\Omega_b h^2$ (blue). The latter is not an informative prior; the BBN prior \cite{Mossa} used in \cite{MM2023} is shown as vertical blue line. The spectral index prior is informative but around five times broader than the Planck primary CMB anisotropy constraint (red vertical line). It saturates the constraint for current noise levels. Broadening the prior can lead to up to a 50\% degradation in the $S_8$ constraint. The actual constraint from \cite{MM2023} and its prior choices are shown as stars.  }
    \label{fig:priordep}
   
\end{figure}

{\bf Degeneracies: } The broad goal for structure growth constraints is to compare their measurement of the amplitude of structure with predictions from the CMB anisotropies. Within flat $\Lambda$CDM, weak lensing observables tend to depend primarily on $\sigma_8$ (or $A_s$), $\Omega_m$, $H_0$ and weakly on $n_s$ and $\Omega_b h^2$. Combinations of $\sigma_8$ and $\Omega_m$ are often reported with $S_8\equiv \sigma_8(\Omega_m/0.3)^{0.5}$ widely used in the galaxy weak lensing literature since this combination roughly captures the amplitude of weak lensing observables at low redshifts.  CMB lensing, however, best constrains the parameter combination $\sigma_8(\Omega_m/0.3)^{0.25}$; generally, probes that derive their information at higher redshifts have a weaker degeneracy with $\Omega_m$ since the universe is more matter dominated and Einstein-de-Sitter-like where $\Omega_m(z)\rightarrow 1$ in a flat cosmology. The degeneracies with $\Omega_m$ and $H_0$ can be partially broken by including BAO data which provides a strong constraint in the $\Omega_m - H_0$ sub-space. As noted in \cite{MM2023},  when attempting to isolate information on $\sigma_8$, CMB lensing benefits significantly more from the addition of BAO data compared to galaxy weak lensing. This is not only due to the weaker dependence on $\Omega_m$ but also because the wide range of scales probed by CMB lensing isolates the constraint into a tube in $\sigma_8-\Omega_m-H_0$ space (through the intersection of two surfaces from the large-scale and small-scale parts of the power spectrum) whose intersection with the BAO surface $\Omega_m-H_0$ is a small localized region. On the other hand, when accounting for the large number of nuisance parameters, galaxy weak lensing primarily probes small scales $k>0.2\impc$ where the power spectrum is approximated by a power law, leading to a sheet-shaped constraint in $\sigma_8-\Omega_m-H_0$ space. The intersection of this constraint with the BAO $\Omega_m-H_0$ surface is not as well localized, leading to a significantly worse $\sigma_8$ constraint.

Similar to previous CMB lensing analyses, the ACT DR6 analysis employed an informative prior on the spectral index $n_s$ as well as a BBN prior \cite{Mossa} on $\Omega_b h^2$. As discussed in \cite{MM2023}, the $n_s$ prior is five times broader than the primary CMB constraint from \Planck~ and therefore a reasonable distillation of CMB anisotropy information on the initial conditions. In Fig. \ref{fig:priordep}, we explore using an information matrix calculation how the width of the $n_s$ and $\Omega_b h^2$ priors affects the relative uncertainty on $S_8$. We see that the width of the $\Omega_b h^2$ prior (e.g. from BBN) has no significant effect on the $S_8$ constraint.  For $n_s$ however, while the prior saturates the constraint for current noise levels, broadening it can lead to up to a 50\% degradation in the $S_8$ constraint.

\subsection{Current status of CMB lensing}

To date, high significance measurements of the CMB lensing auto-spectrum have been made by the ACT \cite{1103.2124,1611.09753,2304.05202}, SPT \cite{1202.0546,1412.4760,1705.00743,1905.05777,2308.11608} and \Planck\cite{1303.5077,1502.01591,1807.06210,2206.07773}~ collaborations with PolarBear \cite{1312.6646} and BICEP \cite{1606.01968}  also having detected the signal.  The most recent and constraining analyses have yielded detection significances of 43$\sigma$, 17$\sigma$ and 42$\sigma$ from ACT \cite{2304.05202}, SPT \cite{2308.11608} and \Planck~ \cite{2206.07773} respectively.  While \Planck~ observations ended in 2013, the collaboration has produced multiple releases incorporating more data and algorithmic improvements. The PR4 lensing analysis \cite{2206.07773} in particular made several improvements to how the input CMB maps are filtered, leading to substantial improvements in SNR. This release has provided a lensing map over 65\% of the sky. The ACT DR6 release \cite{2304.05202,2304.05203} used all data collected by the Advanced ACT receivers from 2017 to 2021 and produced a high-fidelity map for public release. The higher resolution and lower noise levels allowed for a significantly higher fidelity lensing map (in terms of noise per mode) compared \Planck, but the overall SNR of the lensing auto-power-spectrum is comparable to the \Planck~ analysis due to the lower sky area of 23\%.

These measurements of the CMB lensing auto-spectrum offer an important view of the integrated matter density over the epoch of structure formation. Significant effort expended to control systematics at the data vector level have resulted in robust measurements that do not have any nuisance parameters. Moreover, the scales and redshifts probed by CMB lensing are such that linear theory is an excellent (but not perfect) approximation to the models fit, with the choice of non-linear evolution codes and baryonic feedback having negligible impact for current measurements \cite{2011.06582}.

A notable feature of nearly all CMB lensing analyses to date, including the most recent ACT, SPT and \Planck~ releases, is that the amplitude of structure measured from these are in vigorous agreement with the \Planck~ primary CMB anisotropy extrapolation assuming the \LCDM~ model (see Fig. \ref{fig:compilation}).\footnote{This is true with or without the inclusion of BAO data.  Without BAO data (e.g. \cite{2304.05202}), one still obtains a high-precision measurement of $\sigma_8(\Omegam/0.3)^{0.25}$, once again in excellent agreement with the prediction from \Planck.}  As we will see in the next section, this has important consequences for the interpretation of low $S_8$ measurements seen by other growth probes.

\section{Outlook and conclusion}\label{sec:broader}

\begin{figure}
\includegraphics[width=\columnwidth]{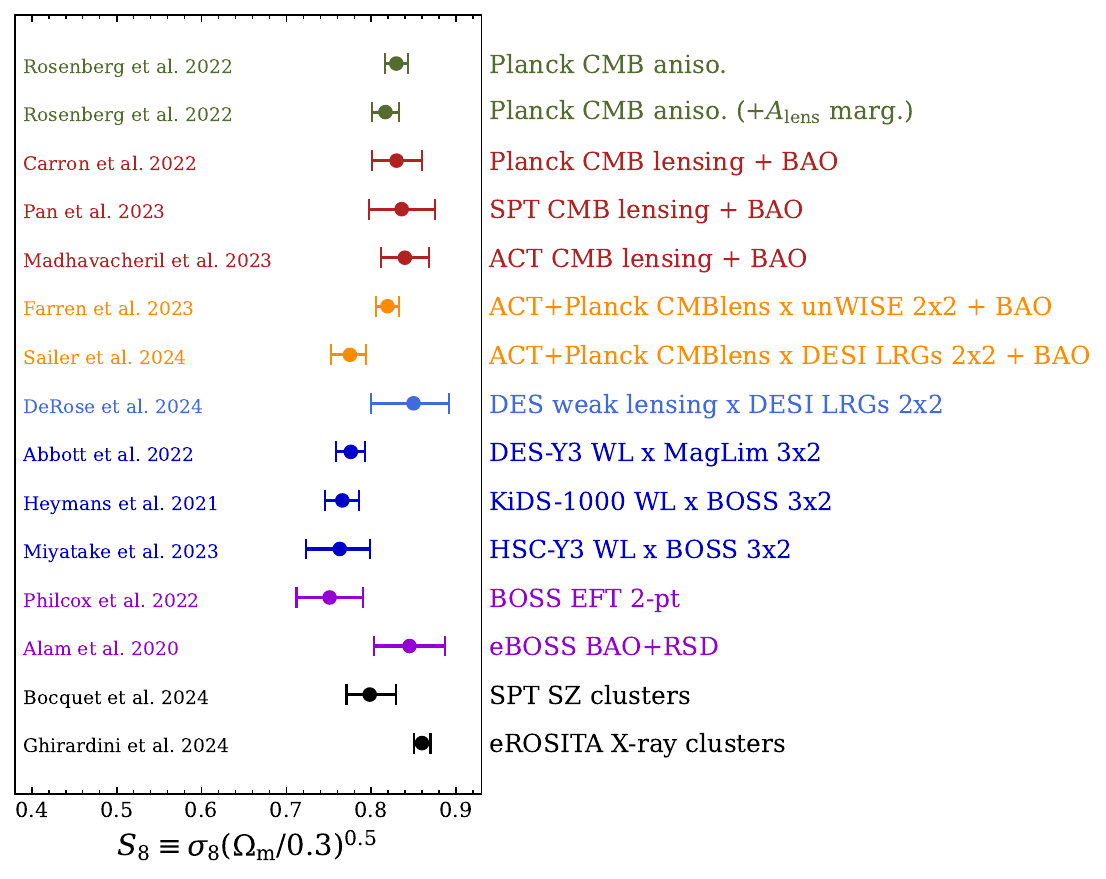}
  \caption{A compilation of $S_8$ constraints from various growth probes compared with primary CMB constraints (green). The priors and parameterizations in these various analyses may not be identical and there may be some covariance between the various growth probes.}
    \label{fig:compilation}
   
\end{figure}

In Fig. \ref{fig:compilation}, we show a compilation of $S_8$ constraints from various growth probes along with primary CMB extrapolations from the \Planck~ experiment (green), with many but not all low-redshift growth probes showing values systematically lower than $S_8=0.8$. Understanding the wave-numbers $k$ and redshifts $z$ probed by various discrepant probes is therefore crucial going forward in distinguishing between new physics and astrophysical systematics. Importantly, as discussed in the previous section, direct probes of intermediate redshifts $z \sim 1-3$ and large scales $k<0.1 \impc$ through the CMB lensing auto-spectrum from \Planck, ACT and SPT are all in excellent agreement with the \Planck~ CMB prediction assuming the \LCDM~ model. There are no other probes of linear scales with comparable precision  at $z>1$ (save for an unWISE CMB lensing cross-correlation discussed later, also consistent with \Planck).   This is a strong indication that if the $S_8$ tension is not due to statistical fluctuations, then its origin must be traced to lower redshifts or higher wave-numbers than the CMB lensing auto-spectrum is significantly sensitive to.\footnote{Measurements of the Lyman-$\alpha$ forest power spectrum from eBOSS sensitive to $k\sim 1 \impc$ and $z=2-5$ (e.g. \cite{2309.03943}) also give low amplitude of structure fits, which is consistent with this interpretation.}

{\bf Primary CMB: } In the standard \LCDM\ parameterization, the \Planck~ CMB anisotropy measurement that uses the temperature and polarization 2-point power spectra (TT, TE and EE) is not a pure primary CMB extrapolation because of the fact that the lensing of the CMB affects the 2-point power spectrum (in addition to inducing mode-coupling and producing a connected 4-point function as discussed earlier).  The effect of lensing is to smear the acoustic peaks and introduce additional small-scale power.  Thus the CMB anisotropies are somewhat sensitive to $S_8$ directly through low-redshift contributions from lensing.  It has been noted \cite{Planck2015} that the amplitude of lensing inferred through the 2-point statistics is higher than that seen through the 4-point information in the CMB lensing reconstruction (though with the PR4 re-processing, this discrepancy has come down significantly \cite{Rosenberg}). Whether this preference for high lensing amplitude is due to a statistical fluctuation or a systematic, one may thus wonder whether the CMB anisotropy $S_8$ `extrapolation' is affected by it. To account for this, one can marginalize over the amplitude $A_{\rm lens}$ of the lensing power spectrum that appears in the expression for the CMB 2-point spectra. The resulting nearly pure-primary CMB extrapolation is also shown; its value is slightly lower, but does not make a qualitative difference to the discussion on $S_8$ discrepancies. Primary CMB measurements by ACT and SPT give similar extrapolations but with larger uncertainties.

{\bf Galaxy weak lensing: }Observables that include cosmic shear auto-correlation information (e.g. HSC \cite{2304.00701,2304.00705,2304.00704}, DES \cite{2105.13549}, KiDS \cite{2007.15632}), whether on their own or in combination with galaxy lensing (in 3x2 analyses), generally tend to show low $S_8$ values each at roughly the $2\sigma$ level, and at the same time have significant weight from small scales $k>0.2 \impc$. Intriguingly, several ``2x2'' analyses that do not include the cosmic shear power spectrum but only the shear-galaxy and galaxy-galaxy correlations agree well with the \Planck~ extrapolation.  Notable among these is the recent DES x DESI analysis \cite{DeRose2024}, where well-characterized DESI spectroscopically-calibrated lens galaxies were used along with effective field theory (EFT) modeling for both non-linear galaxy biasing as well as intrinsic alignments. This EFT modeling is conservative (with many more free parameters than typically used) and consistently retains all parameters that are needed up to a given order in the perturbation theory. The uncertainty on $S_8$ is significantly increased to 6.3\% with the conservative modeling choices but the mean value agrees very well with \Planck. While other past works involving the DES \texttt{MagLim} lens sample have found that non-linear galaxy bias modeling  does not significantly improve agreement with \Planck\cite{2105.13545,2105.13546}, a re-analysis  of this sample in \cite{DeRose2024} (once again, excluding cosmic shear auto-correlations) that uses EFT modeling for intrinsic alignments and stringent scale cuts finds an $S_8$ value that agree well with \Planck. A common feature of such ``2x2'' analyses is that their conservative modeling of systematics shifts the scales they probe to $k<0.2 \impc$.

{\bf CMB lensing cross-correlations: }Cross-correlations with CMB lensing are a powerful way to probe structure formation across a wide range in redshifts while improving control of systematics in galaxy surveys (e.g. \cite{2203.12440,2206.10824}). A particularly useful channel for probing large scales and low redshifts is the ``2x2'' combination of CMB lensing-galaxy correlations together with the galaxy auto-correlation.  This channel excludes galaxy weak lensing altogether but requires careful modeling of non-linear galaxy biasing. If the galaxy sample is spectroscopic (or spectroscopically calibrated), then it has additional robustness against redshift uncertainties. The most high signal-to-noise detection to date of $C_L^{\kappa g}$ is with the unWISE infrared sample in cross-correlation with ACT DR6 CMB lensing \cite{2309.05659}. With only two flux bands, the redshift distribution of this sample has to be calibrated using cross-correlation redshifts \cite{Newman,Menard,Krowleski}. The unWISE x \Planck+ACT analysis leads to a 1.9\% $S_8$ constraint that is in excellent agreement with the \Planck~ extrapolation, with most of its redshift weight at $z>0.6$.  The ACT DR6 CMB lensing map has also recently allowed a cross-correlation analysis with the DESI Luminous Red Galaxies (LRG) sample \cite{2407.04606,2407.04607}. Similar to its galaxy lensing counter-part \cite{DeRose2024}, the lens sample is well-characterized and spectroscopically-calibrated and the analysis uses the same conservative and consistent EFT modeling for non-linear galaxy bias. Unlike \cite{DeRose2024}, no modeling of intrinsic alignments is needed, resulting in a 2.9\% constraint \cite{2407.04606,2407.04607}, albeit with a mean value that is around 2$\sigma$ lower than \Planck.  Since the spectroscopic information allows for a tomographic constraint, this analysis is able to constrain $S_8(z)$ where there is some indication that the lowest redshift bin ($z\approx 0.47$) shows the largest deviation from the \Planck~ prediction (although this could be a statistical fluctuation), inviting further investigation of lower redshifts.

Other probes that measure the growth of structure include (a) galaxy cluster abundances where both SPT SZ-selected clusters and eROSITA X-ray selected clusters find high $S_8>0.8$ \cite{2402.08458} and (b) redshift-space distortions (RSD). For the latter, the eBOSS collaboration analysis \cite{2007.08991} finds an $S_8$ value consistent with \Planck~ but full-shape EFT approaches with the BOSS data (e.g. \cite{2007.08991}) have found values lower than the \Planck~ prediction.  Many other statistics beyond the 2-point function such as the galaxy bispectrum \cite{2112.04515,2406.13388} and higher moments of weak lensing maps \cite{2110.10141} have also consistently been yielding $S_8$ values $2-3\sigma$ lower than \Planck.

{\bf Outlook: } The above results show promising improvements in precision for a wide variety of probes that probe different regions of $P_{mm}(k,z)$. However, astrophysical systematics such as non-linear biasing and intrinsic alignments (requiring many parameters, with associated volume effects due to unconstrained degeneracy directions) are a challenge with current observations for many probes. Unusually large amounts of baryonic feedback (causing suppression in the total matter power spectrum at $k>0.2 \impc$), in particular, has been proposed as a possible resolution for the tension with \Planck~ predictions \cite{10.1051/0004-6361/202346539 ,2206.11794,2305.09827}. It should be noted that this only works for some observables like cosmic shear that probe very small scales, while at the same time measurements of HSC-Y3 cosmic shear seem to be an intriguing exception \cite{2403.20323}.  New measurements of the gas distribution with the kSZ effect \cite{2009.05557,2009.05558,2407.07152} in particular are a promising path forward in breaking the degeneracy of $S_8$ with baryonic feedback.

Meanwhile, intriguing new physics explanations such as late-time modifications of structure growth \cite{2302.01331,2308.16183} and axion dark matter \cite{2301.08361,2311.16377} have been proposed. In his 1993 book surveying the state of physical cosmology, Peebles observed \cite{Peebles1993} 

\begin{quote}
The central issue for this section is whether there is within the standard cosmology a world picture that reconciles the dramatic structures observed on scales of tens of megaparsecs, [...], with the [...] high degree of isotropy of the cosmic radiation backgrounds. If this proved to be impossible, it would show there is something very wrong with the standard model. As will be described, there is no such crisis in theory and observation as now understood...
\end{quote}

Thirty years later, this issue has renewed attention thanks to high precision measurements of the CMB and the plethora of observations of late-time structure being investigated by various large-scale structure and CMB surveys.  The flurry of data expected from the Simons Observatory \cite{1808.07445}, SPT-3G \cite{1407.2973} and CMB-S4 \cite{1610.02743} surveys as well as Euclid \cite{2405.13491}, Roman Space Telescope \cite{10.48550/arXiv.1503.03757}, DESI \cite{1611.00036} and the Rubin Observatory \cite{1809.01669} will allow for a thorough mapping of $P_{mm}(k,z)$. The next decade is thus likely to tell us whether there is something wrong with the standard model, or whether there are one or more astrophysical systematics from the non-linear universe responsible for discrepancies with the CMB growth prediction.

\ack{This review is based on a talk I gave at the Royal Society meeting  "Challenging the standard cosmological model"; I am grateful to the organizers for their invitation to and hospitality during a very fruitful meeting. The review draws heavily from published work from the Atacama Cosmology Telescope. I am grateful to my ACT collaborators for input, including Frank Qu, Gerrit Farren, Blake Sherwin, Boris Bolliet, Neelima Sehgal and Alexander van Engelen. I thank Sebastian Bocquet for providing importance sampled chains for the SPT SZ cluster analysis that fix the sum of neutrino masses. I also thank the two anonymous referees whose feedback significantly improved this manuscript. I acknowledge support from NSF grants AST-2307727 and  AST-2153201 and NASA grant 21-ATP21-0145.}

\appendix

\section{Growth factor at late times}
\label{app:dgrowth}

At late times (well after decoupling), on sub-horizon scales and in linear theory, the evolution of the total matter density is given by the second-order differential equation (see e.g. \cite{Dodelson})

\ba
\frac{d^2 \delta_m}{da^2} + \frac{d {\rm ln} (a^3H)}{da}\frac{d \delta_m}{da} - \frac{3\Omegam H_0^2}{2a^5H^2}\delta_m = 0
\ea

The time-dependence of growth $\delta({\bf k},a) \propto D(a) \delta({\bf k},1)$ can then be obtained by solving this equation numerically.  In certain cases, analytic solutions are possible. With the definition of the growth factor such that $D(a)=1$ in an EdS universe, an integral solution for arbitrary spatial curvature but constant equation of state of dark energy is \cite{1977MNRAS.179..351H}

\ba
D(a) = \frac{5\Omega_m H_0^2}{2} H(a) \int_0^a \frac{da'}{\dot{a'}^{3} }
\ea

A further simplification is possible in a flat universe that ignores radiation at late times, using the Friedman equation $H^2(a) = H_0^2 (\Omegal + a^{-3}\Omegam)$ and defining

$$
x = \left(\frac{\Omegal}{\Omegam}\right)^{\frac{1}{3}} a
$$

we get \cite{astro-ph/1002.3660}

\ba
 D(a) = \frac{5}{2}\left( \frac{\Omega_{\rm M,0}}{\Omega_{\Lambda,0}} \right)^{1/3}\frac{\sqrt{1+x^3}}{x^{3/2}}
                               \int_0^x\frac{x'^{3/2}dx'}{[1+x'^3]^{3/2}}
\ea

The integral above evaluates to

$$
\int_0^x\frac{x'^{3/2}dx'}{[1+x'^3]^{3/2}} = \frac{2}{5}x^{5/2} ~~{}_{2}F_{1}(\frac{5}{6},\frac{3}{2},\frac{11}{6},-x^3)
$$

leading to the final expression

\begin{equation}
\frac{D(a)}{a} = \sqrt{1+x^3} ~~{}_{2}F_{1}(\frac{5}{6},\frac{3}{2},\frac{11}{6},-x^3)
\end{equation}
\begin{equation}
x = \frac{\Omega_\Lambda}{\Omega_m} a
\end{equation}

This expression agrees very well (better than 5\%) at redshifts $z<200$ when compared with numerical solutions of the differential equations in the Boltzmann hierarchy (with the \texttt{CLASS} code), with the disagreement primarily due to ignoring the energy density of radiation. It reduces to $D(a)=a$ when $x\rightarrow 0$ or equivalently $\Omegal\rightarrow 0$.

%%%%%%%%%% Insert bibliography here %%%%%%%%%%%%%%

\bibliography{msm}
\bibliographystyle{RS}

\end{document}